\documentclass[conference]{IEEEtran}
\IEEEoverridecommandlockouts
\usepackage{cite}
\usepackage{amsmath,amssymb,amsfonts}
\usepackage{algorithmic}
\usepackage{graphicx}
\usepackage{textcomp}
\usepackage{multirow}
\usepackage{xcolor}
\usepackage{booktabs}  
\usepackage{dblfloatfix}
\usepackage[table]{xcolor}
\usepackage{natbib}
\usepackage{tabularx}
\usepackage{array}
\usepackage{hyperref}

\definecolor{diagcolor}{RGB}{255,220,220}  
\definecolor{offdiagcolor}{RGB}{245,245,245}  

\def\BibTeX{{\rm B\kern-.05em{\sc i\kern-.025em b}\kern-.08em
    T\kern-.1667em\lower.7ex\hbox{E}\kern-.125emX}}
\begin{document}


\title{\Large \textbf{ACOUSTIC DOMAIN SHIFT IN SPOKEN LANGUAGE IDENTIFICATION \\
FROM SYSTEMATIC DOMAIN GENERALIZATION EVALUATION TO REAL-WORLD APPLICATION}
\thanks{This project was provided with computing HPC and storage resources by GENCI at IDRIS thanks to the grant 2025-AD011017021 on the supercomputer Jean Zay's A100/ H100 partitions.}
}



\author{
\IEEEauthorblockN{François Derrida$^{*}$$^{\dagger}$ \qquad Raphaël Duroselle$^{\dagger}$ \qquad Thomas Courtat$^{*}$ \qquad Jean-François Bonastre$^{\dagger}$}
\IEEEauthorblockA{
$^{*}$THALES, CortAIx-Labs\\
}
\IEEEauthorblockA{
$^{\dagger}$AMIAD Pôle Recherche - Institut Polytechnique de Paris\\
Palaiseau, France \\
francois.derrida@thalesgroup.com
}
}

\maketitle

\begin{abstract}

Domain Generalization (DG) aims to develop models that remain robust to conditions unseen during training. While DG has been systematically studied in computer vision through controlled benchmarks and diverse distribution shifts, its evaluation in spoken language recognition remains less structured. Existing speech datasets provide valuable benchmarks for robustness to real-world acoustic conditions, but are primarily designed around specific scenarios and large scale rather than as general-purpose tools for systematically constructing and evaluating domain shifts. In this work, we introduce a small-scale speech dataset and evaluation protocol for controlled studies of acoustic domain shifts. It enables the evaluation of spoken language recognition models under a variety of acousitc domain shifts. We introduce the speech modality into the DomainBed Domain Generalization framework and evaluate three domain generalization algorithms. We show that in-domain performance is not a reliable predictor of cross-domain robustness and verify that explicit domain invariant algorithms such as MMD or DANN algorithms do not outperform ERM. We further validate the generality of these findings on MMS-LID-126, a state-of-the-art spoken language identification system. We release the code\footnote{\url{https://github.com/AMIAD-Research/DomainBed_extension_audio_adaptation}}. 

\end{abstract}


\section{Introduction}

Deploying machine learning models in real-world settings requires robust Out-of-Distribution (OOD) generalization across domains (i.e., a specific environment, conditions or data distribution), enabling models to transfer from a training \textit{source} domain to an unseen deployment \textit{target} domain \cite{arjovsky2020out}. Yet, most models struggle under substantial distribution shifts \cite{saenko2010adapting, moreno2012unifying} because standard transfer relies on the fragile assumption that \textit{source} and \textit{target} distributions remain close.

Domain generalization is a common challenge for various modalities such as finance \cite{mashrur2020machine}, self-driving cars \cite{yang2025can}, healthcare and medicine \cite{erickson2017machine}. While several datasets address noise robustness in speech \cite{Zhang_2024, leglaive2023chimeUDASE}, they are typically large-scale and involve multiple confounding factors (e.g., speaker, channel, and environment variability), making them less suitable for controlled studies of noise-induced domain shift. This motivates the design of a compact and controlled speech dataset. 

Several studies investigate domain mismatch in Language Identification (LID) and propose domain adaptation strategies, addressing corpus shifts with DANN \cite{abdullah2020cross}, transmission-channel shifts with MMD \cite{duroselle2020unsupervised}, and corruption shifts through flow enhancement \cite{cao2026flow}. Beyond its practical relevance, LID provides a controlled classification setting to study how acoustic domain shifts affect out-of-domain generalization.
Notably,  \cite{dey2024towards} addresses cross-corpora generalization while \cite{muralikrishna2021spoken} considers generalization to an unknown channel.
Nonetheless each of these studies is limited to specific target domain generalization, and less attention has been given to systematically evaluate Domain Generalization under controlled and diverse acoustic domain shifts and across multiple target conditions. This raises the broader question of how to systematically characterize domain shifts, assess their impact on language identification systems, and select appropriate DG strategies accordingly.

Different studies propose to evaluate domain generalization methods \cite{galstyan2021failuremodesdomaingeneralization, cha2021swaddomaingeneralizationseeking}, notably with DomainBed \cite{gulrajani2020searchdomainbed}, which provides a unified benchmark for evaluating algorithms across image datasets, improving reproducibility and comparability. Despite efforts in the speech enhancement literature to cross-validate domains in the multi/mono source settings \cite{Gonzalez_2023}, existing domain generalization benchmarks do not include small speech related datasets. This approach provides a first assessment of robustness under practical deployment conditions. We further evaluate the generalizability of these findings beyond the small-scale benchmark setting using MMS-LID-126, a state-of-the-art spoken language identification system.



In this work, we extend this perspective to speech by investigating domain generalization algorithms under application-relevant acoustic conditions. Here are the main contributions:

\begin{itemize}
    \item We \textbf{show how to build small speech datasets suited to the systematic study of corruption shifts} in classification tasks such as Language Identification (Section \ref{sec:dataset}).
    \item We \textbf{systematically evaluate three Domain Generalization algorithms} and characterize domain generalization under acoustic corruption shifts using the DomainBed experimental protocol (Section~\ref{sec:domainbed}).
    \item We \textbf{validate the generality of our results at large scale} on a state-of-the-art LID system covering 126 languages (Section~\ref{sec:mms}).
\end{itemize}


\section{Speech corruption shift dataset generation}
\label{sec:dataset}

This section presents the proposed methodology to generate a speech dataset named CWWS (Clean, Wham, Wind, Saturation) tailored for a domain generalization classification study in Language Identification (LID). 

\subsection{Corruption shift domains} \label{sec:corruption_shift_domains}
\label{sec:domains}
Following \cite{ben2006analysis}, a domain $\mathcal{D}$ is a probability distribution over $\mathcal{X} \times \mathcal{Y}$, where $\mathcal{X}$ is the space of data (speech utterances) and $\mathcal{Y}$ the set of labels (languages).  Domain generalization is the scenario when the training algorithm has access to data from $N$ source domains $\mathcal{D}_{S_1}, \ldots, \mathcal{D}_{S_N}$, with the goal of generalizing to an unknown and unseen target domain $\mathcal{D}_T$.

We focus on the special case where domains differ solely through corruption shifts. This setting allows us to isolate the impact of acoustic corruptions on out-of-domain generalization. More precisely, we assume there is an underlying original distribution of speech $\mathcal{D}$. All other domains $\mathcal{D}(f)$ are assumed to be defined as $\mathcal{D}(f) = \{ (f(x), y) | (x,y)\sim \mathcal{D}\}$, where $f$ is a potentially stochastic transformation of the speech utterance. This is a simplifying assumption as it is well known that a perturbation of the acoustic conditions impact the production of speech. For instance, speakers systematically modify their vocal production when speaking in noisy environments, a phenomenon known as the Lombard effect \cite{junqua1993lombard}.

We study the following domains, defined by the corruption shift:
\begin{itemize}
    \item \textit{clean}: the original speech data;
    \item \textit{wham}: additive noise with real-world "babble" noise (background conversational speech) taken from the WHAM! dataset \cite{Wichern2019WHAM};
    \item \textit{wind}: additive noise generated using the wind simulator presented in \cite{Mirabiliiwindnoisesimulator2022};
    \item \textit{saturation}: a nonlinear amplitude transformation that induces signal clipping and distortion, with a distortion gain sampled uniformly in [0.5, 4.5];
    \item \textit{demand}: additive noise from DEMAND dataset \cite{thiemann2013demand}; 
    \item \textit{lpc10}: application of the LPC10 speech codec \cite{bittner2016pysox} to the clean audio files;
\end{itemize}

We deal with \textit{clean}, \textit{wham}, \textit{wind}, and \textit{saturation} corruptions in the DomainBed study (Section~\ref{sec:domainbed}) and extend the analysis to the full set of corruptions in the large scale application (Section~\ref{sec:mms}). For additive noises (\textit{wham}, \textit{wind} and \textit{demand}) we consider a range of possible SNRS (uniform distribution over [-10, 10]~dB). 

\subsection{Source speech data}

The corruption-shift dataset is built from an original speech corpus under controlled corruption conditions (see Section~\ref{sec:corruption_shift_domains}). We impose four constraints on the design of the dataset. Firstly, to facilitate comparison with existing works, we respect the train/validation/test splits of the original corpus. Secondly, to evaluate uniquely the corruption shift, the test set of each domain correspond to an augmented version of the exact same list of speech utterances. Third, to compare methods we train all methods with exactly the same number of speech utterances. Finally, to ensure independent sampling across domains, each speaker is assigned to only one domain.

\section{Systematic Analysis of Corruption Domains and adaptation algorithms using LID task }
\label{sec:domainbed}

\begin{table*}[b]
\caption{
DG and Mono-Source (MS) results.
\underline{Underline} values refer to best performance across a given regime and \textbf{bold} values to overall best performance on target domain. Values are accuracies with 95\% bootstrap confidence intervals (2.5th–97.5th percentiles). We highlight \colorbox{diagcolor}{In-Domain} performance in red. AVG refers to the mean across domains.
}
\centering
\small
\setlength{\tabcolsep}{6pt} 
\renewcommand{\arraystretch}{0.9}
\begin{tabular}{l c c c c c}
\toprule
\textbf{Train Domains} & \textbf{clean} & \textbf{wham} & \textbf{wind} & \textbf{saturation} & \textbf{AVG} \\
\midrule
MS clean & \cellcolor{diagcolor} \textbf{\underline{95.3}} {\small[94.6, 95.9]} & 84.8 {\small[83.8, 85.9]} & 73.6 {\small[72.3, 74.8]} & 92.1 {\small[91.4, 92.9]} & 86.5 \\
MS wham & 90.3 {\small[89.4, 91.2]} & \cellcolor{diagcolor} \textbf{\underline{87.4}} {\small[86.4, 88.4]} & 61.1 {\small[59.5, 62.5]} & 85.4 {\small[84.4, 86.4]} & 81.1 \\
MS wind & 93.4 {\small[92.6, 94.0]} & 82.6 {\small[81.5, 83.7]} & \cellcolor{diagcolor} \textbf{\underline{81.2}} {\small[80.1, 82.4]} & 91.9 {\small[91.1, 92.7]} & \textbf{\underline{87.3}} \\
MS saturation & 94.8 {\small[94.1, 95.4]} &  81.3 {\small[80.1, 82.4]} & 70.8 {\small[69.4, 72.1]} & \cellcolor{diagcolor} \underline{92.9} {\small[92.1, 93.7]} & 84.9 \\
\midrule
\cellcolor{diagcolor} MS-In & \cellcolor{diagcolor} \textbf{\underline{95.3}}  &  \cellcolor{diagcolor} \underline{\textbf{87.4}} & \cellcolor{diagcolor} \textbf{\underline{81.2}} & \cellcolor{diagcolor}  \underline{92.9} &\cellcolor{diagcolor} \\ 
Avg MS-OOD & 92.8 & 82.9 & 68.5 & 89.8 & \\

\midrule
\midrule
\textbf{Algorithms} & \textbf{clean} & \textbf{wham} & \textbf{wind} & \textbf{saturation} & \textbf{AVG} \\
\midrule

\multicolumn{6}{c}{\textbf{Domain Generalization - model selection via source domains validation}} \\
ERM-DG  & 93.7 {\small[93.0, 94.4]} & 81.8 {\small[80.7, 82.9]} & \underline{75.2} {\small[73.9, 76.5]} & 92.3 {\small[91.6, 93.0]} & 85.8\\
MMD-DG & \underline{95.1} {\small[94.2, 95.7]} & 81.9 {\small[80.8, 83.0]} & 71.7 {\small[70.3, 73.1]}  & \textbf{\underline{94.6}} {\small[94.0, 95.3]} & 85.8 \\
DANN-DG & 94.8 {\small[94.0, 95.4]} & \underline{83.1} {\small[82.0, 84.2]} & 74.3 {\small[73.0, 75.6]} & 93.6 {\small[92.8, 94.2]} & \underline{86.4} \\
\midrule

\multicolumn{6}{c}{\textbf{Domain Generalization - model selection via  target domain validation (Oracle)}} \\

ERM-DG & 95.0 {\small[94.4, 95.6]} & 82.5 {\small[81.3, 83.5]} & \underline{74.8} {\small[73.5, 76.0]} &  \underline{94.2} {\small[93.5, 94.9]} & \underline{86.6} \\
MMD-DG & \textbf{\underline{95.3}} {\small[94.7, 96.0]} & \underline{82.6} {\small[81.5, 83.7]} & 71.9 {\small[70.6, 73.3]} &  94.0 {\small[93.3, 94.7]} & 85.9 \\
DANN-DG & 94.3 {\small[93.6, 94.9]} & 82.3 {\small[80.1, 82.4]} & 74.6 {\small[73.4, 75.8]} & 93.4 {\small[92.6, 94.0]} & 86.2 \\
\bottomrule
\end{tabular}
\label{tab:All_absolute_results}
\end{table*}

As demonstrated by \cite{gulrajani2020searchdomainbed}, evaluating domain generalization algorithms requires a controlled design where each algorithm is evaluated with several pairs of source and target domains and with the same budget of hyperparameter search. This section presents how we use the CWWS dataset to perform the speech corruption-shift evaluation.

\subsection{Evaluation with DomainBed}

The evaluation is performed with the DomainBed framework \cite{gulrajani2020searchdomainbed} which embeds numerous domain generalization algorithms. Assuming cross-domain invariance objective, we treat the multiple source domains as a single aggregated source, thereby reducing the multi-source setting to a binary source–target adaptation problem. In this paper we focus on the following three algorithms, that have been successfully applied to LID \cite{duroselle2020unsupervised,abdullah2020cross}:

\begin{itemize}
    \item \textit{Empirical Risk Minimization (ERM)} \cite{vapnik1999overview} seeks to minimize the overall prediction error by optimizing the cumulative loss across all domains and data points
    \item \textit{Maximum Mean Discrepancy (MMD)} \cite{sun2016deep, gretton2012kernel} minimizes the discrepancy between the means of two distributions in a high-dimensional feature space induced by a kernel. Regularization is achieved by aligning all pairs of source domains
    \item \textit{Domain-Adversarial Neural Network (DANN)} \cite{ganin2016domain} employs an adversarial objective to align feature distributions across domains. Regularization is performed for all source domains at once.
\end{itemize}

\noindent \textit{Keep-one-out evaluation:} Each algorithm is evaluated on four target domains: \textit{clean}, \textit{wham}, \textit{wind} and \textit{saturation}. For each experiment we keep one domain out (target domain) and use the remaining ones as source domains. We report results in the four possible configurations. We compare these DG  configurations with mono-source (MS) experiments for which only one domain is seen during training.

\noindent \textit{Model selection methods:} One important aspect of domain generalization evaluation is the fair selection of hyperparameters for each algorithm and set of domains. We perform a grid search with 20 experiments for each configuration. It results in 80 runs per considered algorithm. From this set of experiments, we report performance of models selected using the source-domain validation set and also those selected using target-domain validation set (Oracle).

\noindent \textit{Uncertainties:} Contrary to the original framework \cite{gulrajani2020searchdomainbed}, we do not report uncertainties due to the random seed. Instead, we measure the statistical uncertainty due to the size of the test set with 95\% bootstrap confidence intervals.

\subsection{Data and model architecture}
\label{sec:data_arch}

The number of classes and utterances follows canonical DG benchmarks such as VLCS \cite{Fang_2013_ICCV}. To ensure a comparable training budget, all configurations use 4,500 training utterances in total. The DG setting distributes these utterances equally across three domains (1,500 utterances per domain), whereas the mono-source setting uses all 4,500 utterances from a single domain. Table~\ref{tab:domainbed_dataset_summary} synthesizes the corresponding statistics.



\begin{table}[htbp]
\centering
\caption{Dataset characteristics. DG and MS training runs use 4,500 samples. "Total utterances'' denotes the total number of utterances in the dataset, and Utterances/domain'' their distribution across domains. All sets are language-balanced.}
\label{tab:domainbed_dataset_summary}

\begin{tabularx}{\columnwidth}{@{}l
    >{\centering\arraybackslash}X
    >{\centering\arraybackslash}X
    >{\centering\arraybackslash}X
    >{\centering\arraybackslash}X@{}}
\hline
& \multicolumn{2}{c}{\textbf{Train}}
& \textbf{Validation}
& \textbf{Test} \\
\cmidrule(lr){2-3}
\textbf{Split used in}
& DG & MS & DG \& MS & DG \& MS \\
\hline
Total utterances       & 6,000 & 18,000 & 6,000 & 18,000 \\
Utterances/domain      & 1,500 & 4,500 & 1500 & 4,500 \\
$\neq$ Speakers     & 2,389 & 2,389 & 1,500 & 870 \\
\hline
\end{tabularx}
\end{table}

Inspired by URGENT Challenge \cite{Zhang_2024} LID downstream task, we use the original speech corpus CommonVoice \cite{ardila2020commonvoice} (v22) with the four following languages: French, Spanish, German, and Mandarin Chinese.

The architecture of the model is similar to \cite{pratap2024scaling}. It is a pretrained wav2vec2 speech encoder \cite{baevski2020wav2vec2}, followed by mean pooling and a classification head. We use a backbone\footnote{https://huggingface.co/facebook/wav2vec2-large-100k-voxpopuli} implemented in the transformer library \cite{wolf2020transformers} and pretrained on VoxPopuli \cite{wang2021voxpopuli}, a large-scale multilingual corpus with diverse speakers and recording conditions, disjoint from CommonVoice to reduce the risk of potential information leakage \cite{yu2024rethinking}. Experiments show that backbone-only pretraining yields near-random classification, making fine-tuning essential for the considered shifts; results are omitted for brevity.

\begin{table*}[h!]
\centering
\caption{Test accuracy of MMS-LID-126 systems (\%) on Fleurs (FL) an VoxLingua107 (VL) corpus. \underline{Underline} values refer to best performance across a given regime and \textbf{bold} values to overall best performance on target domain. We highlight \colorbox{diagcolor}{In-Domain} performance in red.  
}
\small
\setlength{\tabcolsep}{6pt} 
\renewcommand{\arraystretch}{1}
\begin{tabular}{lcccccc|cccccc}
\toprule
 \textbf{Trainings  Domains} & \multicolumn{2}{c}{\textbf{clean}} & \multicolumn{2}{c}{\textbf{wham}} & \multicolumn{2}{c}{\textbf{saturation}} & \multicolumn{2}{|c}{\textbf{wind}}  & \multicolumn{2}{c}{\textbf{lpc10}} & \multicolumn{2}{c}{\textbf{demand}}\\
 
Corpus & FL & VL & FL & VL & FL & VL & FL & VL & FL & VL & FL & VL   \\
\midrule
mms-lid-126 & \cellcolor{diagcolor} \textbf{\underline{96.4}}  & \cellcolor{diagcolor} 94.0 & 79.1 & 80.7 & 90.1 & 84.9 & \textbf{\underline{96.1}} & 93.7 & 71.4 & 69.7 & 84.2 & 83.8 \\
\hline
MS-clean & \cellcolor{diagcolor} 95.6 & \cellcolor{diagcolor} 93.5 & 76.9 &  79.1 & 86.0 & 83.7 & 95.2 & 93.0 & 61.8 & 66.9 & 82.5 & 83.0 \\
MS-wham & 93.7 & 93.8 & \cellcolor{diagcolor} \textbf{\underline{87.6}} & \cellcolor{diagcolor} \textbf{\underline{87.1}} &  85.6 &  82.8 & 93.5 & 93.7 &  68.1 & 70.5 & 88.2 & 87.5 \\
MS-saturation & 95.4 & 92.9 & 71.2 & 73.5 & \cellcolor{diagcolor} \textbf{\underline{94.2}} & \cellcolor{diagcolor} \textbf{\underline{92.6}} &  94.3 &  91.7 & 63.8 & 62.5 & 78.9 & 78.7 \\
MS-wind & 95.6 & 93.8 & 77.3 & 78.7 & 86.1 & 81.4 & \cellcolor{diagcolor} 95.7 & \cellcolor{diagcolor} 93.6 &   62.3 & 67.1 & 83.3 & 82.2  \\
MS-lpc10 & 93.9 & 93.7 & 68.0 & 71.3 & 79.0 & 75.8 & 93.3 & 92.9 & \cellcolor{diagcolor} \textbf{\underline{92.2}} & \cellcolor{diagcolor} \textbf{\underline{89.7}} & 75.5 &  76.3  \\
MS-demand & 94.7 & \underline{94.1} & 85.7 & 84.7 & 87.4 & 83.6 & 94.9 & \textbf{\underline{94.4}} & 71.1 & 71.8 & \cellcolor{diagcolor} \textbf{\underline{89.4}}  & \cellcolor{diagcolor} \textbf{\underline{88.9}} \\
\hline
\cellcolor{diagcolor} MS-In & \cellcolor{diagcolor} \textbf{\underline{96.4}} & \cellcolor{diagcolor} 94.0 & \cellcolor{diagcolor} \textbf{\underline{87.6}} & \cellcolor{diagcolor} \textbf{\underline{87.1}} & \cellcolor{diagcolor} \textbf{\underline{94.2}} & \cellcolor{diagcolor} \textbf{\underline{92.6}} & \cellcolor{diagcolor} 95.7 & \cellcolor{diagcolor} 93.6& \cellcolor{diagcolor} \textbf{\underline{92.2}} & \cellcolor{diagcolor} \textbf{\underline{89.7}}& \cellcolor{diagcolor} \textbf{\underline{89.4}} & \cellcolor{diagcolor} \textbf{\underline{88.9}} \\
Avg MS-OOD & 94.7 & 93.7 & 76.4 & 78. & 85.7 & 82.0 & 94.6 & 92.2 & 66.4 & 68.1 & 82.1 & 81.9\\
\hline
\hline
 & \multicolumn{12}{c}{\textbf{Domain Generalization experiments}} \\
\hline
\textbf{Algorithms} & \multicolumn{6}{c}{ \cellcolor{diagcolor} \textbf{In-Domain}} & \multicolumn{6}{|c}{\textbf{Out-of-Domain}} \\
\hline
ERM-DG & \cellcolor{diagcolor} 95.1 & \cellcolor{diagcolor} \textbf{\underline{94.4}} & \cellcolor{diagcolor} 87.4 & \cellcolor{diagcolor} \underline{86.8} & \cellcolor{diagcolor} 93.7 & \cellcolor{diagcolor} 92.3 & \underline{95.2} & \underline{94.0} &  68.7 & 68.1 & 88.9 & \underline{88.1}  \\
MMD-DG & \cellcolor{diagcolor} 94.8 & \cellcolor{diagcolor} 93.9 & \cellcolor{diagcolor} \underline{87.5} & \cellcolor{diagcolor} 86.2 & \cellcolor{diagcolor} 93.8 & \cellcolor{diagcolor} \underline{92.4} & 94.9 & 93.7 & \underline{70.1} & \underline{69.2} & 88.9 & 87.3 \\
DANN-DG & \cellcolor{diagcolor} \underline{95.2} & \cellcolor{diagcolor} 93.4 & \cellcolor{diagcolor} \underline{87.5} & \cellcolor{diagcolor} 85.2 & \cellcolor{diagcolor} \underline{93.9} & \cellcolor{diagcolor} 92.2 & \underline{95.2} & 93.3 &  68.6 & 66.3 & \underline{89.0} & 87.1 \\
\bottomrule
\end{tabular}
\label{tab:mms}
\end{table*}

\subsection{Results} \label{sec:domainbed_results}



The first part of Table~\ref{tab:All_absolute_results} reports the Mono-Source (MS) experiments. Diagonal entries correspond to in-domain evaluations (highlighted in red), while off-diagonal entries measure the impact of domain shifts. As expected, we illustrate the domain-shift problem since in-domain performance outperform out-of-domain performance and sometimes by a large margin (\textit{wind} domain). Also, in-domain results show that domains are not equally difficult with a relatively easy \textit{clean} domain, \textit{wind} much more harder and \textit{wham} somewhere in between. Interestingly, domain "difficulty" and domain "usefulness" for generalization  are different concepts: \textit{wind} is difficult in-domain but is useful as a source. OOD performance show that generalization is asymmetric. For instance \textit{wham} leads to quite poor performance on \textit{wind} (61.1\%) when \textit{wind} is relatively good as a source domain to perform on \textit{wham} (82.6\%) despite they are both additive corruptions.

The second part of Table~\ref{tab:All_absolute_results} shows that multi-source Domain Generalization (DG) outperforms average Out-of-Domain performance obtained in mono-source settings and for some domains the In-domain performance (\textit{saturation}). Surprisingly, the best OOD performance from MS-clean on \textit{wham} domain outperforms DG performance. As concluded by \cite{gulrajani2020searchdomainbed} for image datasets, the simple ERM algorithm does not seem to be outperformed by MMD or DANN algorithms. It shows that explicit invariance objectives do not consistently improve generalization abilities. Two model selection methods are reported and we observe that Oracle selection adds little. Target-domain information provides only marginal model-selection gains.

These observations motivate the question of whether the same behavior persists in a realistic large-scale LID setting.

\section{Large-Scale Robust Language Identification}
\label{sec:mms}

Although comparable in terms of number of languages to recent domain adaptation studies \cite{duroselle2020unsupervised,cao2026flow}, this controlled experimental setting does not reflect the scale of modern language identification systems that can handle hundreds to thousands languages \cite{jia2022compact,pratap2024scaling}. We extend our analysis to a reference language identification system: \textit{mms-lid-126\footnote{https://huggingface.co/facebook/mms-lid-126}}  \cite{pratap2024scaling}.

This set of experiments aims both at validating the results of the previous section on a large scale system, and at producing a robust language recognition system. The hyperparameters of the LID training recipe are taken from \cite{pratap2024scaling} whereas the hyperparameters of the domain generalization algorithms are selected from the previous section.

\subsection{Domain generalization with mms-lid-126}

The architecture of the mms-lid-126 model \cite{pratap2024scaling} is identical to Section~\ref{sec:data_arch}, replacing the pretrained wav2vec2 encoder with the mms-1b model\footnote{https://huggingface.co/facebook/mms-1b}. MMS stands for \textit{Massively MultiLingual Speech} and corresponds to pretraing on a very diverse corpus with more than 44.7K hours and 126 languages.

Following \cite{pratap2024scaling}, the models are trained on the union of FLEURS \cite{conneau2023fleurs} and VoxLingua107 \cite{valk2021voxlingua107}, resulting into 126 languages.  FLEURS (audio books) and VoxLingua107 (more noisy recordings from Youtube) correspond to different speech styles and acoustic conditions. 

Performance is evaluated on the six domains defined in section \ref{sec:domains}. Domain generalization experiments are performed with the three following source domains: \textit{clean}, \textit{wind} and \textit{saturation}. The three remaining (\textit{wind}, \textit{lpc10} and \textit{demand}) corruptions are considered as unseen during finetuning.

\subsection{Results}

First we observe an important performance drop of the pretrained mms-lid-126 system for all corruption shifts. Our MS-clean system achieves a similar performance on the clean domain but is less robust to corruption shifts. We hypothesize that the mms-lid-126 system is more robust because it has been trained on more data (MMS-lab-U+unlab+FLEURS+VoxLingua107 in \cite{pratap2024scaling}).

The qualitative observations from previous analysis (Section~\ref{sec:domainbed_results}) persist with a much larger language inventory. The same relationship between source-domain performance and transferability is observed. \textit{Wham} domain remains a relatively transferable source particularly for close domains such as \textit{demand}. This further demonstrates that our conclusions generalize across corpora. The relative behavior of ERM, MMD, and DANN is broadly consistent across FLEURS and VoxLingua107, despite differences in absolute performance.

Finally, results on LPC10 corruption show that domains diversity help but is not always sufficient to guarantee robustness to arbitrary unseen corruptions.

\section{Conclusion}

In this work, we introduced a method for building speech datasets to systematically evaluate Domain Generalization under corruption shifts. By addressing the lack of systematic DG evaluation in speech, we extend the DomainBed framework to the speech modality, providing the community with a simple and reproducible way to benchmark and compare DG methods. Our experiments suggest that robustness to acoustic domain shifts in language identification is driven primarily by exposure to diverse acoustic conditions rather than by enforcing explicit domain invariance. However, this robustness is strongly dependent on the source-target relationship: corruption transfer is asymmetric, and difficult unseen distortions can remain challenging even after multi-domain training. We expect this protocol to facilitate further research on speech domain generalization. Beyond corruption shifts, speech domains also vary in accents, emotions, and speaking styles, often intertwined with acoustic conditions and transmission channels. A natural next step is to extend this framework from zero-shot domain generalization to few-shot domain adaptation.







\newpage

\newpage
\small
\bibliographystyle{unsrt}
\bibliography{biblio}

\end{document}